\documentclass[12pt]{article}
\usepackage[colorlinks=true,linkcolor=blue, linktoc=page, urlcolor=blue,citecolor=magenta]{hyperref}
\usepackage{amsmath,amscd}
\usepackage{amsfonts}
\usepackage{amssymb}
\usepackage{graphicx,shortvrb}
\usepackage{epsfig}
\usepackage{newlfont}
\usepackage{cite}
\usepackage[utf8]{inputenc}

\def\calk{\cal K}
\def\call{\cal L}

\def\calo{\cal O}
\def\calt{\cal T}

\def\b{\beta}

\def\G{\Gamma}

\def\l{\lambda}

\def\k{\kappa}

\def\m{\mu}
\def\n{\nu}

\def\r{\rho}

\def\s{\sigma}

\newcommand{\os}[2]{{\overset{\,\scalebox{0.5}{(#2)}}{#1}}{}}

\begin{document}

\begin{titlepage}
 \vskip 1.8 cm

\begin{center}{\huge \bf
Torsional Newton-Cartan gravity from the large $c$ expansion of General Relativity} 
\end{center}
\vskip .3cm
\vskip 1.5cm

\centerline{\large {{\bf Dieter Van den Bleeken}}}

\vskip 1.0cm

\begin{center}
\sl Primary address\\
 Physics Department, Boğaziçi University\\
 34342 Bebek / Istanbul, TURKEY

\vskip 1cm

Secondary address\\
Institute for Theoretical Physics, KU Leuven\\
3001 Leuven, Belgium

\vskip 1cm

\texttt{dieter.van@boun.edu.tr}

\end{center}

\vskip 1.3cm \centerline{\bf Abstract} \vskip 0.2cm \noindent

We revisit the manifestly covariant large $c$ expansion of General Relativity, $c$ being the speed of light. Assuming the relativistic connection has no pole in $c^{-2}$, this expansion is known to reproduce Newton-Cartan gravity and a covariant version of Post-Newtonian corrections to it. We show that relaxing this assumption leads to the inclusion of twistless torsion in the effective non-relativistic theory. We argue that the resulting TTNC theory is an effective description of a non-relativistic regime of General Relativity that extends Newtonian physics by including strong gravitational time dilation.

\end{titlepage}

\tableofcontents
\section{Introduction}
In our everyday life most\footnote{GPS technology provides an exception, as it has recently become a part of many peoples day to day life and it crucially relies on knowledge of gravitational time dilation, an effect not captured by Newtonian physics.} gravitational phenomena are perfectly well described by Newtonian physics. But by making very precise measurements one can observe deviations from this theory, which are predicted by Einsteins theory of General Relativity (GR). The series of deviations is captured in the so called Post-Newtonian expansion, see \cite{Poisson:2014} for a pedagogic introduction. It is a valid approximation to GR in a regime of weak gravitational fields and small velocities. In most applications, where one is interested in computing these corrections as efficiently as possible, one starts by choosing a coordinate system that simplifies the relevant fields as much as possible. From a more conceptual point of view one could wonder if this expansion can be performed without giving up the general covariance of GR. Indeed this is the case, and it can be shown \cite{Kunzle:1976, Ehlers:1981, Dautcourt:1990} that the leading order in such a covariant expansion produces so called Newton-Cartan gravity, a covariant formulation of Newtonian gravity. Furthermore this covariant procedure can be extended to higher order \cite{Dautcourt:1996pm, Tichy:2011te} and coincides with the more standard Post-Newtonian expansion once particular coordinates are chosen.

In this paper we revisit the non-relativistic expansion of \cite{Dautcourt:1990, Dautcourt:1996pm, Tichy:2011te}, which is formulated as a large $c$ expansion of a family of metrics parametrized by the speed of light $c$. In that derivation it is assumed that the relativistic metric is such that the associated Levi-Cevita connection remains finite in the large $c$ limit. Although this might appear a natural assumption at first, one should keep in mind that the metric is allowed to diverge as $c\rightarrow \infty$. So why not the connection one could ask. In this work we relax this assumption and find that it leads to rather interesting observations, both mathematically and physically. 

From the mathematical point of view we show that first of all everything, including the geodesic equation, remains consistent if one allows the Levi-Civita connection to formally diverge.  Moreover the effective theory that describes approximate solutions to the Einstein equations up to Next to Next to Leading order (NNLO) in large $c$ is a generalization of standard Newton-Cartan theory where the Newton-Cartan connection now contains torsion. Interestingly enough such torsional Newton-Cartan geometries have been studied only recently in a rather different set of contexts, see for example \cite{Christensen:2013lma, Bergshoeff:2014uea, Bekaert:2014bwa, Hartong:2015zia, Geracie:2015dea, Banerjee:2016laq, Bergshoeff:2017btm}. The theory that appears out of the large $c$ expansion of GR has, as a consequence of the Einstein equations, what is called in the previously mentioned literature {\it twistless} torsion. It is standard to refer to this particular case of Newton-Cartan geometry as TTNC geometry. Although the geometry is rather well understood, the possible dynamic equations for its fields, turning the theory into something that could be called TTNC gravity, seems to be not very much explored. In \cite{Afshar:2015aku} equations where constructed using conformal tensor calculus methods, but the equations we find here, see table \ref{eomtable}, appear different and are as far as we know only the second set of explicit equations compatible with TTNC geometry that have been found so far. In \cite{Hartong:2015zia} a set of actions in 3 dimensions was constructed and a relation between TTNC gravity and Hořava-Lifshitz gravity was established. It would be interesting to see if there is some overlap with our effective theory arising out of GR.

From the physical point of view our work appears interesting as the torsion corresponds to a non-trivial timelike warpfactor already at the leading $c^2$ order. This should be compared to the standard Newtonian potential which appears in the same warpfactor, but at the subleading $c^0$ order. This implies first of all that this expansion captures effects that are {\it not} included in the standard Post-Newtonian expansion, as that expansion starts from the assumption of gravitational fields weak enough so that the $c^2$ timelike warpfactor is trivial, or in other words it assumes the metric is a weak field correction to the Minkowski metric. The generalized expansion we work out here also works around metrics which do not necessarily have this weak field form. In particular, the additional non-relativistic torsion that it includes appears at a {\it lower} order than the Newtonian potential, or said differently, at 'Pre-Newtonian' order. In summary it seems our expansion captures certain non-relativistic, but strong gravitational physics that is absent in the standard Post-Newtonian expansion which only captures non-relativistic, weak gravitational effects.

Although the previous arguments are nice conceptually, it would be extremely interesting if one could find actual real world physical situations that fall into this non-relativistic strong gravity regime, and to see if this generalized expansion can have any practical use.

Finally let us stress that although our effective non-relativistic theory has torsion, it is an approximation at large $c$ of standard relativistic GR {\it without} torsion.

The paper is organized as follows. First we rewrite the large $c$ expansion of the relativistic metric in terms of variables naturally appearing in Newton-Cartan geometry in section \ref{secgeom}. In section \ref{eqsec} we  then compute the non-relativistic equations that are equivalent to the relativistic Einstein equations up to NNLO in large $c$. The remaining part of the paper is then used to discuss various mathematical and physical aspects of the effective non-relativistic theory, in section \ref{comsec}. Note that we there also provide an explicit example. Finally we have added appendix \ref{partsec}, where we discuss the geodesic equation in this expansion, to show that allowing the relativistic connection to diverge at large $c$ does not lead to inconsistencies. Although we provide a number of technical details there we feel a better physical understanding of the expanded geodesic equations is still missing.

\section{Expansion of the geometry}\label{secgeom}
In this section we introduce the expansion of the relativistic metric in powers of $c$, following \cite{Dautcourt:1990, Dautcourt:1996pm, Tichy:2011te}. We then repackage the independent coefficients in this expansion into objects that appear naturally in the formalism of non-relativistic geometry, see for example \cite{Hartong:2015zia}. As the standard Poisson equation of Newtonian gravity appears at NNLO in the equations of motion \cite{Dautcourt:1990}, we will keep track of all independent fields that appear up to NNLO. A number of those fields happen to drop out of the equations of motion up to NNLO, but we want to stress and make explicitly clear that this is an outcome, not an assumption. For this reason we work out the expansion of the geometry without imposing any dynamical constraints yet. Furthermore in \cite{Dautcourt:1990, Dautcourt:1996pm, Tichy:2011te} it was assumed that as $c\rightarrow \infty$ the relativistic Levi-Civita connection remains finite. We will perform the expansion without making this assumption and we will discuss the physical interpretation and consequences of this in section \ref{schsec}.

\subsection{Starting ansatz}
We assume an expansion of the metric ($D=d+1$, Lorentzian) in even powers\footnote{One can argue  \cite{Dautcourt:1990} that odd powers of $c$ will only appear at higher order than we are interested in. This implies that one does not lose any generality by restricting to even powers here. Still this will be one of the few assumptions we put into the formalism from the start. It might be interesting to allow odd terms in the expansion from the beginning and see directly from the equations of motion that they can be consistently put to zero.} of a variable $c$ (thought of physically as the speed of light):
\begin{equation}
g_{\m \n} = \sum_{i=-1}^\infty \os{g}{2$i$}\!_{\mu\nu} c^{-2i} \qquad g^{\m \n}= \sum_{i=0}^\infty\os{g}{2$i$}^{\m\n} c^{-2i}\label{metexp}
\end{equation}
We furthermore assume that $\os{g}{-2}\!_{\mu\nu}$ is of rank 1 and negative, so we can write
\begin{equation}
\os{g}{-2}\!_{\mu\nu}=-\tau_\mu\tau_\nu
\end{equation}

\subsection{Diffeomorphisms}\label{difsec}
Before we start a detailed analysis of this expansion and its consequences it will be useful to investigate its behavior under diffeomorphisms, as was stressed in \cite{Tichy:2011te}. Where general relativity is invariant under coordinate transformations that can be arbitrary functions of $c$, the ansatz \eqref{metexp} is only preserved by diffeomorphisms that are analytic in $c^{-2}$. These are generated by vectorfields of the form
\begin{equation}
\xi^\mu=\sum_{i=0}^\infty \os{\xi}{$2i$}^\mu c^{-2i}
\end{equation}
For tensors of the form $T=\sum_{i=i_\mathrm{min}}^\infty \os{T}{$2i$}c^{-2i}$ the coefficients then transform as
\begin{equation}
\delta_\xi \os{T}{$2i$}=\call_{\os{\xi}{0}}\os{T}{$2i$}+\sum_{j=i_\mathrm{min}}^{i-1}\call_{\os{\xi}{$2i-2j$}}\os{T}{$2j$}
\end{equation}
First of all we see that all tensor coefficients transform as tensors under $c$-independent diffeomorphisms, generated by  $\os{\xi}{0}$. For this reason we will from now on simply refer to the transformations generated by $\os{\xi}{0}$ as 'the' diffeomorphisms of the expansion. 

In addition there is an infinite amount of additional gauge symmetries, that originate from the diffeorphisms with subleading $c$-dependence, which act in a more non-trivial way, mixing different coefficients. Note that up to some fixed order of the expansion there is however only a finite number of those that act non-trivially, as under those transformations a given coefficient only gets contributions from lower order coefficients, never from higher order ones.

It will be useful to repeat this same analysis in the case of a connection. Assuming a connection with the expansion $\G^\l_{\m\n}=\sum_{i=-1}^\infty\os{\G}{$2i$}_{\m\n}^\l c^{-2i}$ one computes that
\begin{equation}
\delta_{\xi}\os{\G}{$2i$}_{\m\n}^\l=\call_{\os{\xi}{0}}\os{\G}{$2i$}_{\m\n}^\l+\partial_\mu\partial_\nu \os{\xi}{$2i$}^\lambda+\sum_{j=-1}^{i-1}\call_{\os{\xi}{$2i-2j$}}\os{\G}{$2j$}_{\m\n}^\l\label{condif}
\end{equation}
This transformation is interesting as it tells us that only one coefficient of the connection will act as a connection in the expanded theory while all others will behave as tensors. Indeed, as we identified $\os{\xi}{0}^\mu$ as the diffeomorphisms we see that it is only $\os{\G}{0}_{\m\n}^\l$ which transforms as a connection, while all other coefficients transform as tensors under $\os{\xi}{0}^\mu$ generated diffeomorphisms. Again there are the additional symmetries generated by the $\os{\xi}{$2i$}^\mu$, $i>0$, under which the coefficients transform in a more complicated fashion.

\subsection{Metric invertibility}
The two expansions \eqref{metexp} are of course related by the condition that one series provides the inverse of the other. We can expand this condition $g_{\m\r}g^{\r\n}=\delta_{\mu}^{\nu}$ order by order and solve the resulting equations explicitely. As a first step one obtains from the leading equation (order $c^2$) the result that
\begin{equation}
\os{g}{0}^{\mu\nu}=h^{\m\n}\qquad\mbox{with}\qquad h^{\m\n}\tau_\nu=0\label{hdef}
\end{equation}  
For the actual metric $g_{\m\n}$ to be non-degenerate $h^{\m\n}$ will need to have rank 3.

Before continuing it turns out to be useful to introduce two new, but dependent, fields $\tau^\mu$ and $h_{\m\n}$ that are defined via the conditions
\begin{equation}
\tau_\nu\tau^\nu+h_{\m\r}h^{\r\n}=\delta^\n_\m\qquad \tau^\r\tau^\s h_{\r\s}=0\label{projdef}
\end{equation}
These equations uniquely define the new fields only if we make an identification by gauge transformations of the form
\begin{equation}
\delta_\chi \tau^\mu=-h^{\m\r}\chi_\r\qquad \delta_\chi{h_{\m\n}}=\tau_\m\chi_\n+\tau_\n\chi_\m\qquad \label{boosttransfo}
\end{equation}
Here the gauge parameter $\chi_\m$ is purely 'spatial', i.e. $\tau^\r\chi_\r=0$.
This gauge transformation which we here introduced 'by hand' corresponds to the local Galilean boost symmetry in the Newton-Cartan literature, see e.g. \cite{Andringa:2010it}, so we will henceforth also refer to such transformations as boost transformations.

The main use of the relation \eqref{projdef} is that it provides two complementary projectors
\begin{equation}
\tau_\m{}^\n=\tau_\mu\tau^\nu \qquad h_\m{}^\n=h_{\m\r}h^{\r\n}
\end{equation}

We can now decompose all further metric coefficients in the expansions \eqref{metexp} along these projectors and this is a great help in solving the expanded inverse condition. The result is that both metric and inverse metric coefficients up to NNLO can be written in terms of two independent vector fields $C_\m$ and $B_\m$ and two symmetric 'spatial' tensors $\beta^{\m\n}$ and $\gamma^{\m\n}$. More precisely one finds that
\begin{eqnarray*}
\os{g}{0}\!_{\m\n}&=&2\tau_{(\mu}C_{\n)}+h_{\m\n}\\
\os{g}{2}^{\m\n}&=&- \tau^\mu\tau^\n+2\tau^{(\mu}h^{\nu)\l}C_\l+\beta^{\m\n}\\
\os{g}{2}\!_{\m\n}&=&B_\mu\tau_\nu+\tau_\m B_\n-C_\m C_\n-h_{\m\r}h_{\n\s}\beta^{\r\s}\\
\os{g}{4}^{\m\n}&=&\left(h^{\r\s}C_\r C_\s-2\tau^\r C_\r\right)\tau^\mu\tau^\nu+2\tau^{(\mu}h^{\n)\r}\left( B_{\r}+(C_\s\tau^\s)C_\r+h_{\r\l}C_\s \b^{\s\l}\right)+\gamma^{\m\n}
\end{eqnarray*}
where $\tau_\r\beta^{\r\m}=\tau_\r\gamma^{\r\m}=0$ and $\beta^{[\m\n]}=\gamma^{[\m\n]}=0$.

The metric coefficients are of course defined independently of the projectors we introduced and so should not transform under the boost symmetry \eqref{boosttransfo}. This then implies that the new fields need to transform as\footnote{It is straightforward to work out the boost transformations of the fields $B_\mu$ and $\gamma^{\mu\nu}$. As they will play no role in the rest of this work we don't explicitely write out these transformations here.}
\begin{equation}
\delta_\chi C_\mu=-\chi_\mu\qquad\qquad
\delta_\chi \b^{\mu\n}=-2\chi^{(\m} h^{\n)\r}C_\r
\end{equation}

Because the original metric $g_{\m\n}(c)$ is invariant under the boost transformations all derived objects and physical equations should be expressible in terms of boost-invariant quantities. There is a natural set of those:
\begin{eqnarray}
\hat \tau^\m&=&\tau^\m-h^{\m\n}C_\n\label{htau}\\ \hat{h}_{\m\n}&=&h_{\m\n}+2\tau_{(\m}C_{\n)}+2 \hat{\Phi}\tau_\m\tau_\n\\ \hat{\Phi}&=&-\tau^\r C_\r+\frac{1}{2}h^{\r\s}C_\r C_\s\\
\hat{\beta}^{\m\n}&=&\beta^{\m\n}+h^{\m\r}h^{\m\s}C_\r C_\s\label{hbeta}\\
\hat{B}_\m&=&B_\m+h_{\m\r}\beta^{\r\s}C_\s+\frac{1}{2}\tau_\m \left(\beta^{\r\s}C_\r C_\s+(\tau^\r C_\r-h^{\r\s}C_\r C_\s)^2\right)\nonumber\\
&&-C_\m(\tau^\r C_\r-h^{\r\s}C_\r C_\s)\\
\hat{\gamma}^{\m\n}&=&\gamma^{\m\n}+2h^{\r(\m}h^{\n)\s}C_\r\left(\hat B_\s-2\hat{\Phi}C_\s\right)
\end{eqnarray}
Note that these hatted, boost invariant variables satisfy constraints similar to the orginal fields:
\begin{equation}
\tau_\nu\hat \tau^\nu+\hat h_{\m\r}h^{\r\n}=\delta^\n_\m\qquad \hat\tau^\r\hat\tau^\s\hat h_{\r\s}=0\qquad \tau_\m \hat{\beta}^{\m\n}=0\qquad \tau_\m\hat\gamma^{\m\n}=0\label{hprojdef}
\end{equation}
It will also be useful to define the boost invariant projectors
\begin{equation}
\hat{\tau}_{\m}{}^\n=\tau_\m \hat \tau^\n\qquad \hat{h}_{\m}{}^\n=\hat h_{\m\r} h^{\r\n}
\end{equation}

We have summarized the expression of the coefficients of the metric and its inverse up to NNLO in terms of these boost invariant variables in table \ref{metrictable}.

\begin{table}
\framebox[1.1\linewidth]{\begin{minipage}{\linewidth}

\vspace{0.1cm}

\begin{center}
\bf{\caption{\label{metrictable}The metric and its inverse.}}
\end{center}
\paragraph{LO}
\begin{eqnarray*}
\os{g}{-2}\!_{\m\n}&=&-\tau_\mu\tau_\nu\\
\os{g}{0}^{\m\n}&=&h^{\m\n}
\end{eqnarray*}
\paragraph{NLO}
\begin{eqnarray*}
\os{g}{0}\!_{\m\n}&=&-2\hat \Phi \tau_\m\tau_\n+\hat{h}_{\m\n}\\
\os{g}{2}^{\m\n}&=&- \hat\tau^\mu\hat\tau^\n+\hat\beta^{\m\n}
\end{eqnarray*}
\paragraph{NNLO}
\begin{eqnarray*}
\os{g}{2}\!_{\m\n}&=&\tau_\m\hat B_\nu+\tau_\n\hat B_\mu-\hat{h}_{\m\r}\hat{h}_{\n\s}\hat{\beta}^{\r\s}\\
\os{g}{4}^{\m\n}&=& 2\hat{\Phi}\hat{\tau}^\m\hat{\tau}^\nu+2\hat{\tau}^{(\mu}h^{\n)\r}\hat{B}_\r+\hat{\gamma}^{\m\n}
\end{eqnarray*}

\vspace{0.1cm}

\end{minipage}}
\end{table}

\subsection{Metric compatibility}
The conditions that $\nabla_\m g_{\n\r}=0$ and $\nabla_\m g^{\n\r}=0$ with respect to the Levi-Civita connection can also be expanded order by order. At LO this leads to a trivial algebraic identity if one uses the explicit form of the metric coefficients obtained by the inverse condition, see table \ref{metrictable}. More interesting is the NLO part of the above compatibility conditions which read 
\begin{eqnarray*}
\os{\nabla}{0}_\mu h^{\nu\l}&=&-\os{\G}{-2}_{\m\r}^\n \os{g}{2}^{\r\l}-\os{\G}{-2}_{\m\r}^\l \os{g}{2}^{\r\n}\\
\os{\nabla}{0}_\mu\left(\tau_\nu\tau_\l\right)&=&-\os{\G}{-2}_{\m\n}^\r \os{g}{0}\!_{\r\l}-\os{\G}{-2}_{\m\l}^\r \os{g}{0}\!_{\r\n}
\end{eqnarray*}
Let us first note that indeed these are good tensorial equations, as we learned from the analysis in section \ref{difsec} that indeed $\os{\G}{0}_{\m\n}^\l$ transforms as a connection while $\os{\G}{-2}_{\m\n}^\l$ transforms as a tensor. The above equations suggest that when $\os{\G}{-2}_{\m\n}^\l\neq0$, the connection $\os{\G}{0}_{\m\n}^\l$ is not the most natural connection for the expanded theory, as it is not compatible with the structure provided by $h^{\m\n}$ and $\tau_\m$. This is the first place where our work differs from that of \cite{Dautcourt:1990, Dautcourt:1996pm, Tichy:2011te}, where $\os{\Gamma}{-2}_{\m\n}^\l$ was {\it assumed} to be zero.  Working out the explicit form of the RHS of the above equation in terms of the fields appearing in the metric coefficients one realizes however that the above equations can equivalently be rewritten as
\begin{equation}
\os{\nabla}{nc}_\mu h^{\nu\l}=0\qquad\mbox{and}\qquad
\os{\nabla}{nc}_\mu\tau_\nu=0\label{nccompat}
\end{equation}
where
\begin{equation}
\os{\G}{nc}_{\m\n}^\l=\os{\G}{0}_{\m\n}^\l+\left(-\hat{\tau}^\l\hat{\tau}^\r+\hat{\beta}^{\l\r}\right)\left(\tau_\m\partial_{[\n}\tau_{\r]}+\tau_\n\partial_{[\m}\tau_{\r]}-\tau_\r\partial_{[\m}\tau_{\n]}\right)\label{ncdef}
\end{equation}
We will refer to this connection as the Newton-Cartan connection. Working out the explicit form of $\os{\G}{0}_{\m\n}^\l$ in terms of the metric coefficients one finds, via table \ref{metrictable}, that
\begin{eqnarray}\label{ncexplicit}
\os{\G}{nc}_{\m\n}^\l&=&\frac{1}{2}h^{\l\r}\left(\partial_\m h_{\r\n}+\partial_\n h_{\m\r}-\partial_\r h_{\m\n}\right)+h^{\l\r}\tau_{(\m}K_{\n)\r}+\tau^\l\partial_{\m}\tau_\n\\&&+h^{\l\r}\left(C_\m\partial_{[\n}\tau_{\r]} +C_\n\partial_{[\m}\tau_{\r]}-C_\r\partial_{[\m}\tau_{\n]}\right)\nonumber\\
&=&\frac{1}{2}h^{\l\r}\left(\partial_\m \hat h_{\r\n}+\partial_\n \hat h_{\m\r}-\partial_\r \hat h_{\m\n}+2\partial_\r \hat \Phi \tau_\m\tau_\n-4\hat{\Phi}(\tau_\m\partial_{[\n}\tau_{\r]}+\tau_\n\partial_{[\m}\tau_{\r]})\right)+\hat\tau^\l\partial_{\m}\tau_\n\nonumber
\end{eqnarray}
where
\begin{equation}
K_{\m\n}=\partial_\m C_\n-\partial_\n C_\m\,.
\end{equation}
First of all it is interesting to note that contrary to $\os{\Gamma}{0}_{\m\n}^\l$ the connection $\os{\G}{nc}_{\m\n}^\l$ is independent of the field $\hat{\beta}^{\m\n}$. Furthermore from the second expression for $\os{\G}{nc}_{\m\n}^\l$ in \eqref{ncexplicit} one sees that it is manifestly boost-invariant and that it has a torsion $\os{\mathrm{T}}{nc}^\l{}_{\m\n}=2\os{\G}{nc}_{[\m\n]}^\l$, which itself is also boost-invariant, that is given by
\begin{eqnarray}
\os{\mathrm{T}}{nc}^\l{}_{\m\n}&=&2\hat\tau^\l\partial_{[\m}\tau_{\n]}\,.\label{torsion}
\end{eqnarray}

The compatibility conditions \eqref{nccompat} together with the degeneracy condition \eqref{hdef} are the defining equations of Newton-Cartan, or Galilean, geometry see for example \cite{Bekaert:2014bwa}. Although traditionally the connection is furthermore assumed to be torsionless the more general case including torsion has been introduced and studied in for example \cite{Hartong:2015zia, Bekaert:2014bwa}. What is interesting is that our analysis gives a precise meaning to all the fields appearing in the Newton-Cartan connection \eqref{ncdef} in terms of particular coefficients in the expansion of a relativistic, $c$ dependent metric, through table \ref{metrictable} and equation \eqref{metexp}.

\section{Expansion of the Einstein equations}\label{eqsec}
In this section we feed the expansion \eqref{metexp}, rewritten as in table \ref{metrictable},  into the Einstein equations up to NNLO. The resulting equations, summarized in table \ref{eomtable}, provide a dynamics for the fields of the torsional Newton-Cartan geometry that appeared in the previous section. When one puts the torsion to zero the results reduce to that of \cite{Dautcourt:1990, Dautcourt:1996pm, Tichy:2011te}.

\subsection{Setup}
We will write the Einstein equations as
\begin{equation}
R_{\m\n}=8\pi G_\mathrm{N}\,\calt_{\m\n}\,,\qquad \calt_{\m\n}=c^{-4}\left(T_{\m\n}-\frac{1}{D-2}g_{\m\n}g^{\r\s}T_{\r\s}\right)
\end{equation}
Given the expansion of the metric \eqref{metexp} one finds a corresponding expansion of the Ricci tensor
\begin{equation}
R_{\m\n}=\sum_{i=-2}^{\infty}\os{R}{2$i$}_{\mu\nu} c^{-2i}\label{ricexp}
\end{equation}
For this to be consistent with the Einstein equations also the energy momentum must have a similar expansion:
\begin{equation}
\calt_{\m\n}=\sum_{i=-2}^{\infty}\os{\calt}{2$i$}_{\mu\nu} c^{-2i}\label{texp}
\end{equation}
Although in full generality one could consider $\os{\calt}{-4}_{\m\n}\neq 0$ we will restrict ourselves in this work to the case where
\begin{equation}
\os{\calt}{-4}_{\m\n}=0\qquad\qquad \mbox{(assumption)}\label{emassump}
\end{equation}

\subsection{LO: Twistless torsion}
The coefficients in this expansion can now be explicitly computed in terms of the fields appearing in table \ref{metrictable}. For the leading term of the Ricci tensor one finds
\begin{eqnarray}
\os{R}{-4}_{\m\n}&=&\tau_\mu\tau_\nu h^{\k\l}h^{\r\s}\partial_{[\k}\tau_{\r]}\partial_{[\l}\tau_{\s]}\label{ricconstr}
\end{eqnarray}
Together with the assumption \eqref{emassump} this leads to the LO equation in table \ref{eomtable}.

It is instructive to provide some details on the solution of that LO equation. The first of the relations \eqref{hprojdef} implies we can make the following decomposition
\begin{equation}
\partial_{[\mu}\tau_{\n]}=2\tau_{[\m}\hat a_{\n]}+\hat h_{\m\r}\hat h_{\n\s}\hat a^{\r\s}
\end{equation}
where $\hat a^{\m\n}$ is antisymmetric and without loss of generality  we can choose
\begin{eqnarray}
\hat \tau^\r \hat a_\r&=&0\\
\tau_\r \hat a^{\r\m}&=&0
\end{eqnarray} Using this decomposition it then immediatly follows that the LO equation in table \ref{eomtable} is equivalent to $\hat a^{\m\n}=0$. We can then summarize 
\begin{equation}
h^{\k\l}h^{\r\s}\partial_{[\k}\tau_{\r]}\partial_{[\l}\tau_{\s]}=0\qquad\Leftrightarrow\qquad\partial_{[\mu}\tau_{\n]}=\tau_{[\m}\hat a_{\n]}\qquad\Leftrightarrow\qquad \tau_{[\mu}\partial_\n\tau_{\l]}=0\label{adef}
\end{equation}
Here the implication in the last equivalence is straightforward and the other direction follows from the observation that for any $p$-form $\omega$ and 1-form $\tau$ such that $\omega\wedge\tau=0$ there needs to exist a $p-1$-form $\alpha$ such that $\omega=\tau\wedge \alpha$. 

Note that the 1-form $\hat a_\mu$ is not a new independent field, but is directly related to the 1-form $\tau$ via
\begin{equation}
\hat a_\mu=\call_{\hat \tau} \tau_\mu\,.
\end{equation}
This form also explicitely shows that $\hat a_\mu$ is boost invariant.

The condition \eqref{adef} can be seen as constraints on the torsion \eqref{torsion} of the Newton-Cartan connection. Torsion satisfying this constraint has appeared before in the Newton-Cartan literature, see for example \cite{Hartong:2015zia, Bekaert:2014bwa} and is referred to as {\it twistless torsion}. It has the physical interpretation of providing a spacelike foliation, so that all observers can agree on a common direction of time. This condition for the torsion to be twistless simplifies the expansion at higher orders quite a bit, and it will be used in the rest of this paper. As this condition is directly related to the assumption \eqref{emassump} one could contemplate relaxing it, which would further generalize the expansion. We leave this for possible future work.

\subsection{NLO \& NNLO: TTNC gravity}
Using the twistlessness of the torsion obtained at LO we now extend the calculation to NLO and NNLO. Through straightforward but somewhat tedious algebra the relevant coefficients of the Ricci tensor are computed to be
\begin{eqnarray}
\os{R}{-2}_{\m\n}\!&=&\!-\tau_\mu\tau_{\n}h^{\l\r}D_{\l}\hat a_\r\label{higherRic}\\
\os{R}{0}_{\m\n}\!
&=&\!\os{R}{nc}_{\m\n}+\hat h_\m{}^\r \hat{h}_{\n}{}^\s D_{\r} \hat{a}_\s+\tau_\m\tau_\n\!\left(h^{\r\s}\hat{a}_\r \partial_\s(\hat{\Phi}+\frac{1}{2}\hat{\beta})- D_\r(\hat{\beta}^{\r\s}\hat{a}_\s)\right)+\overline{\cal K}_{\m\r}\tau_\n h^{\r\s} \hat{a}_\s\nonumber
\end{eqnarray}
Here $\os{R}{nc}_{\m\n}$ is the Ricci tensor of the connection \eqref{ncexplicit}. To keep the expressions compact we introduced the additional notation:
\begin{eqnarray}
\hat{\beta}=\hat{\beta}^{\r\s}\hat h_{\r\s}\,,\qquad&
D_\mu=\os{\nabla}{nc}_\m-\hat a_\m&\qquad\mbox{and}\quad 
\overline{\cal K}_{\m\n}=\hat{\calk}_{\m\n}-\hat{h}_{\m\n}h^{\r\s}\hat{\calk}_{\r\s}\,.
\end{eqnarray}
The following extrinsic curvature appears:
\begin{equation}
\hat\calk_{\m\n}
=\frac{1}{2}\call_{\hat\tau} \hat h_{\m\n}=\hat{h}_{\r(\m}\hat{h}_{\n)}{}^{\s}\os{\nabla}{nc}_\s \hat{\tau^\r}
\end{equation}
The equations of motion follow by equating the Ricci coefficients to the corresponding energy momentum and can be found in table \ref{eomtable}. Note that consistency requires
\begin{equation}
h^{\m\n}\os{\calt}{-2}_{\n\r}=0\,.
\end{equation}

\begin{table}
\framebox[1.1\linewidth]{\begin{minipage}{\linewidth}

\vspace{0.1cm}

\begin{center}
\bf{\caption{\label{eomtable}The equations of motion.}}
\end{center}
\paragraph{LO}
\begin{equation*}
h^{\k\l}h^{\r\s}\partial_{[\k}\tau_{\r]}\partial_{[\l}\tau_{\s]}=0
\end{equation*}
\paragraph{NLO}
\begin{equation*}
-\tau_\mu\tau_\nu h^{\l\r}D_{\l}\hat a_\r=8\pi G_\mathrm{N}\os{\calt}{-2}_{\m\n}
\end{equation*}
\paragraph{NNLO}
\begin{equation*}
\os{R}{nc}_{\m\n}=-\hat h_\m{}^\r \hat{h}_{\n}{}^\s D_{\r} \hat{a}_\s-\tau_\m\tau_\n\left(h^{\r\s}\hat{a}_\r \partial_\s(\hat{\Phi}+\frac{1}{2}\hat{\beta})- D_\r(\hat{\beta}^{\r\s}\hat{a}_\s)\right)-\overline{\cal K}_{\m\r}\tau_\n h^{\r\s} \hat{a}_\s+8\pi G_\mathrm{N}\os{\calt}{0}_{\m\n}
\end{equation*}
\vspace{0.1cm}

\end{minipage}}
\end{table}

\section{Comments on the effective TTNC gravity theory}\label{comsec}
In this section we collect a number of remarks about the system of equations in table \ref{eomtable}, which for simplicity we will refer to as the TTNC equations. 

\subsection{Gauge Symmetries}
The TTNC equations of motion in table \eqref{eomtable} are invariant under a number of symmetries. The most obvious one, due to the manifest tensorial form of the equations, is that under diffeomorphisms. As explained in section \ref{difsec}, these diffeomorphisms of the effective theory have their origin in the $c$ independent diffeomorphisms $\os{\xi}{0}$ of General Relativity. 

A second symmetry which is rather manifest is the boost symmetry \eqref{boosttransfo}, since only boost invariant objects appear in the TTNC equations. 

There is however a third gauge symmetry, which is less manifest. It originates in the fact that the original expansion up to NNLO is also invariant under diffeomorphisms that are proportional to $c^{-2}$, generated by the $\os{\xi}{2}$ of section \ref{difsec}. Using the decomposition in table \ref{metrictable} we can express the action of these subleading diffeos on the metric as transformations of the non-relativistic effective fields. In doing so it is useful to introduce the notation
\begin{equation}
\Lambda=\tau_\r \os{\xi}{2}^\r\,,\quad \zeta^\m=\hat{h}_\r{}^\m \os{\xi}{2}^\r\,,\qquad D_\m\Lambda=\partial_\m\Lambda+\hat{a}_\m\Lambda
\end{equation}
One then finds the following transformations\footnote{Fields not appearing, like $\tau_\mu$ and $h^{\m\n}$, are invariant.}
\begin{eqnarray}
&\delta_\Lambda\hat{\tau}^\mu=h^{\m\r}D_\r\Lambda \qquad
\delta_\Lambda\hat{h}_{\mu\n}=-2\tau_{(\m}\hat{h}_{\n)}{}^\r D_\r\Lambda
\qquad\delta_\Lambda\hat{\Phi}=\hat{\tau}^\r D_\r\Lambda&\nonumber\\
&\delta_\Lambda\hat{a}_\mu=-\tau_\m \hat{a}_\r h^{\r\s}D_\s\Lambda\qquad\qquad
\delta_\Lambda\hat{\beta}^{\mu\nu}=-2 h^{\m\r}h^{\n\s}\hat{\calk}_{\r\s}\Lambda&
\end{eqnarray}
and
\begin{equation}
\delta_\zeta\hat{\Phi}=-\hat{a}_\r \zeta^\r\qquad\qquad
\delta_\zeta\hat{\beta}^{\mu\nu}=-2h^{\r(\m}\os{\nabla}{nc}_\r \zeta^{\n)}
\end{equation}
Note that as any tensor, see section \ref{difsec}, also the energy momentum coefficients transform:
\begin{equation}
\delta_\Lambda \os{\calt}{0}_{\m\n}=\call_{\Lambda\hat{\tau}}\os{\calt}{-2}_{\m\n}\qquad\qquad \delta_\zeta \os{\calt}{0}_{\m\n}=\call_{\zeta}\os{\calt}{-2}_{\m\n} 
\end{equation}
Although somewhat involved, one can explicitly check that the TTNC equations of table \ref{eomtable} are invariant under these local transformations.

\subsection{Absence and removal of higher order fields}
Note that when expanding out the metric and its inverse up to NNLO, as in table \ref{metrictable}, they are composed out of a rather large set of different objects: 
\begin{equation}
\tau_\mu, h^{\m\n}, \hat \tau^\m, \hat h_{\m\n}, \hat{\Phi}, \hat{\beta}^{\m\n}, \hat B_\mu, \hat{\gamma}^{\mu\nu}\,.
\end{equation}
The first observation is that the fields $\hat B_\mu$ and $\hat{\gamma}^{\m\n}$ do not appear at all in the equations of motion up to NNLO, see table \ref{eomtable}. This implies that at this order of approximation they are completely undetermined and one would need to go to higher order to fix them.

At first sight the field $\hat{\beta}^{\m\n}$ appears more mysteriously. It enters the equations of motion in table \ref{eomtable}, in a purely algebraic way, but is absent in other discussions of TTNC geometry in the literature. The key is to realize it can always be removed by using the extra gauge transformations, discussed in the previous subsection, that are also absent in other discussions of TTNC. Indeed, $\hat{\beta}^{\m\nu}$ only appears in the equations of motion through $\hat{\beta}$ and $\hat{\beta}^{\m\n}\hat{a}_\n$, which constitute $D$ arbitrary functions. But these also transform non-trivially under exactly $D$ independent gauge transformations, explicitly 
\begin{equation}
\delta\hat{\beta}=-2(\os{\nabla}{nc}_\r\zeta^\r+\Lambda\hat{\calk})\qquad\qquad \delta(\hat{\beta}^{\mu\nu}\hat{a}_\n)=-2(h^{\r(\m}\os{\nabla}{nc}_\r \zeta^{\n)}+ h^{\m\r}h^{\n\s}\hat{\calk}_{\r\s}\Lambda)\hat{a}_\n
\end{equation}
So one can always go to a gauge where $\hat{\beta}=\hat{\beta}^{\m\n}\hat{a}_\n=0$ {}\footnote{Note that also the energy momentum might depend on $\hat{\beta}^{\m\n}$, in that case the gauge that would make $\hat{\beta}^{\m\n}$ disappear could be slightly different, but we expect the argument to still hold.}, as this would amount to solving a set of $D$ linear (first-order differential) equations for $\Lambda$ and $\zeta$. 

After this gauge fixing we find ourselves exactly in the established framework of TTNC, with fields $\tau_\mu, h^{\m\n}, \hat \tau^\m, \hat h_{\m\n}, \hat{\Phi}$ and no extra gauge transformations. Note that up to boost transformations these fields are all determined in terms of $\tau_\mu$, $h^{\m\n}$ and $C_\mu$, and that our TTNC equations provide exactly the right number of equations to solve for all unconstrained components of those.

From the point of view of the relativistic metric \eqref{metexp} that we are approximating, we have learned that certain components of the NLO and NNLO coefficients can be put to zero by $c$ dependent coordinate transformations, other components are left completely arbitrary and can only be fixed by going to higher order, while a few components, such as for example $\hat{\tau}^\m\hat{\tau}^\n\os{g}{-4}_{\m\nu}=\hat{\Phi}$ are completely determined already by the equations of motion at NNLO.

\subsection{Relation to other non-relativistic gravity theories}
First we comment on what happens when we put the torsion $\hat a_\mu$ to zero. It is important to note that one is only free to do so in case $\os{\calt}{-2}_{\m\n}=0$, otherwise the torsion will be sourced and one is forced to turn it on. It is easy to check that when the torsion is put to zero the equations reduce to that of standard torsionless Newton-Cartan gravity. This is of course no surprise as in this case our analysis reduces to that of \cite{Dautcourt:1990, Dautcourt:1996pm, Tichy:2011te}. One interesting observation is however that in the torsionless case the fields $\hat{\beta}^{\m\n}$ drop automatically out of the equations in table \ref{eomtable}. This means we no longer need to fix the additional gauge transformations to remove this field. The $\zeta$ transformations are there but act trivially on all the fields that enter in the equations up to NNLO. The $\Lambda$ transformation is more interesting. Exactly when the torsion is zero it simplifies to a U(1) action on the field $C_\mu$:
\begin{equation}
\delta_\Lambda C_\mu=-\partial_\mu\Lambda
\end{equation}
This is the well known U(1) symmetry related to the Bargmann central extension of the Galilei algebra \cite{Andringa:2010it} and so we see that it arises out of GR as a diffeomorphism subleading in $c$.

It is interesting to wonder if our TTNC theory of table \ref{eomtable} is related to other theories recently proposed. In \cite{Afshar:2015aku} dynamical equations for TTNC were constructed, but a detailed comparison reveals that the equation determining  the Ricci tensor of the NC-connection is different. Furthermore, although in that reference the torsion is twistless, they do not impose an additional equation for the torsion, as our NLO equation in table \ref{eomtable} does. It appears that contrary to torsionless NC, for TTNC there are many inequivalent equations of motion consistent with all the symmetries. One of those appears out of an expansion of GR and it would be interesting to understand if it has some special features compared to the others, or if there is a different relativistic origin for these others as well. 

It is possible that our equations are an example of a Hořava-Lifshitz (HL) gravity. The relation between TTNC dynamics and HL gravity was worked out in full generality in three dimensions in \cite{Hartong:2015zia}. This was done at the level of the action however and no equations of motion were provided. It would be interesting to compare this to our TTNC equations by either computing the equations of motion following from \cite{Hartong:2015zia} or by deriving an action for our equations.

\subsection{Physical interpretation and an example solution}\label{schsec}
The direct origin in GR allows us to give an interpretation of the physical role torsion plays in the effective non-relativistic theory in table \ref{eomtable}. When there is no torsion, i.e. $d\tau=0$, we can always choose coordinates such that $\tau_\mu=\delta_\m^0$. Via the expansion \eqref{metexp} and table \ref{metrictable} we see that the leading timelike warpfactor, at order $c^2$, is trivially 1. When there is non-zero twistless torsion, i.e. $\tau\wedge d\tau=0$, we can choose coordinates such that $\tau_\mu= e^\frac{\psi}{2}\delta_\m^0$. In the relativistic metric \eqref{metexp} this implies that there is now a non-trivial timelike warpfactor\footnote{Some readers might wonder if one can not always go to Gaussian normal coordinates and put the timelike warpfactor to 1. This is true, but it would not respect the form of the expansion \eqref{metexp}. In particular it would introduce fractional powers of $c$, as can be seen for example in the case of the Schwarzschild metric in Gaussian normal coordinates: 
\begin{equation*}
ds^2=-c^2d\sigma^2+\frac{2G_\mathrm{N}m}{c^2r(\sigma,\rho)}d\rho^2+r(\sigma,\r)^2d\Omega^2 \qquad r(\sigma,\rho)=\left(\frac{3}{2}\sqrt{2G_\mathrm{N}m}(\sigma+\r c^{-1})\right)^{2/3}\,.
\end{equation*}

In case the theory under consideration is not pure gravity, but also contains a dilaton-like scalar, then one can go to a different conformal frame where the metric has a trivial leading timelike warpfactor, see for example \cite{Julia:1994bs, Bleeken:2015ykr}.} $e^\psi$ at order $c^2$. Note that the NLO equation in table \ref{eomtable} provides an effective nonrelativistic equation of motion for this warpfactor $e^\psi$. In summary we can draw the conclusion that this generalized expansion also includes strong time dilation effects, which are absent in the standard post-Newtonian expansion. As these strong time dilation appears at a {\it lower} order than the Newtonian potential, one could refer to the equations LO a NLO in table \ref{eomtable} as 'pre-Newtonian' order. 

Note that the existence of this generalized expansion should not be too much of a surprise. Indeed, as discussed in detail in e.g. \cite{Poisson:2014}, the Post-Newtonian expansion is not simply a large $c$ expansion but also a weak field expansion. In particular it describes the non-relativistic sector of the so called post-Minkowskian expansion which starts from metrics with small deviations from Minkowski space. This however begs the question if there could be strong field regimes where the physics remains non-relativistic. The expansion of this paper seems to describe exactly such a regime. In case the torsion is non-vanishing we have an order $c^2$ timelike warpfactor, meaning that at leading order the metric is no longer Minkowski but rather some background with strong gravity. Said in yet another way, the torsion seems to include strong time dilation effects which in the standard Newtonian regime are absent. Our expansion shows they can be included while preserving the non-relativistic character of the effective theory.

It will be illustrative to analyze these remarks in an explicit example. Apart from providing some insight in the physical regime described by the effective TTNC gravity in table \ref{eomtable} this will also provide an explicit solution to those equations, providing a check on the mathematical consistency of our expansion.

Let us quickly review the standard Newtonian description of the Schwarzschild metric in our formalism:
\begin{eqnarray}
ds^2&=&-c^2\left(1-\frac{2mG_\mathrm{N}}{rc^2}\right)dt^2+\left(1-\frac{2mG_\mathrm{N}}{rc^2}\right)^{-1}dr^2+r^2d\Omega^2\label{weakbhmetric}\\
&=&-c^2 dt^2+\frac{2mG_\mathrm{N}}{r}dt^2+dr^2+r^2d\Omega^2+\calo(c^{-2})
\end{eqnarray}
Via the expansion \eqref{metexp} and table \ref{metrictable} one reads of that this metric corresponds to the nonrelativistic fields
\begin{equation}
\tau_0=1\,,\ \ \tau_i=0\,,\qquad h_{ij}=\delta_{ij}\,,\ \ h_{0\mu}=0\,,\qquad C_0=\frac{m G_\mathrm{N}}{r}\,,\ \ C_i=0\,.
\end{equation}
In particular we see that this is a torsionless configuration, $d\tau=0$, and we recognize the Newtonian potential of a point mass in $C_0$, it is well known that this solves the NC equations to which the TTNC equations in table \ref{eomtable} reduce in this case. All of this illustrates some comments made above, as we see that in the large $c$ limit the non-trivial part of the warp factor is subleading and the metric at leading order is Minkowski space, so we are not only in the non-relativistic regime but also in the weak field regime.  

This last remark however suggests a way to probe another regime of the Schwarzschild metric. In the previous expansion we formally send $c\rightarrow \infty$ while keeping $m$ fixed. We could consider scaling the mass such that we keep $M=m/c^2$ fixed. The real world analog, where $c$ is large but not infinite, of this would be a situation where $\frac{2mG_\mathrm{N}}{rc^2}=\calo(1)$. In this regime the expansion of Schwarzschild looks rather different:
\begin{equation}
ds^2=-c^2\left(1-\frac{2MG_\mathrm{N}}{r}\right)dt^2+\left(1-\frac{2MG_\mathrm{N}}{r}\right)^{-1}dr^2+r^2(d\theta^2+\sin^2\theta d\phi)\label{bhmetric}
\end{equation}
Note that the formal expansion \eqref{metexp} of this metric actually truncates at order $c^0$. We can now simply read off that
\begin{eqnarray}
\tau_\mu&=&\left(1-\frac{2MG_\mathrm{N}}{r}\right)^{1/2}\delta_{\m}^t\\
h_{\mu\nu}&=&\begin{pmatrix}
0&0&0&0\\
0&\left(1-\frac{2MG_\mathrm{N}}{r}\right)^{-1}&0&0\\
0&0&r^2&0\\
0&0&0& r^2\sin^2\theta\\
\end{pmatrix}\\
C_\m&=& B_\mu=0\\
\beta_{\m\n}&=&\psi^{\m\n}=0
\end{eqnarray}
From this one computes that
\begin{equation}
\hat a_\mu=\left(0,-\frac{MG_\mathrm{N}}{r^2}\left(1-\frac{2MG_\mathrm{N}}{r}\right)^{-1},0,0\right)
\end{equation}
So we now have an effective non-relativistic description with non-zero torsion! Doing a few further computations one can check that the non-relativistic fields above do indeed provide a non-trivial solution of the TTNC equations in table \ref{eomtable}. A few useful results in that computation are that $\hat{\calk}_{\m\n}=0$ and
\begin{eqnarray*}
\os{R}{nc}_{\m\n}&=&-\hat h_\m{}^\r \hat{h}_{\n}{}^\s D_{\r} \hat{a}_\s=\frac{M}{r^3}\hat{h}_{\m\n}-3\frac{r}{MG_\mathrm{N}}\left(1-\frac{2MG_\mathrm{N}}{r}\right)\hat a_\m\hat a_\n\\
&=&\frac{MG_\mathrm{N}}{r^3}\begin{pmatrix}
0&0&0&0\\
0&-\left(1-\frac{2MG_\mathrm{N}}{r}\right)^{-1}&0&0\\
0&0&r^2&0\\
0&0&0& r^2\sin^2\theta\\
\end{pmatrix}
\end{eqnarray*}

This second way of expanding the Schwarschild\footnote{One can consider a similar expansion for the Kerr metric \cite{vdbproc}. } metric illustrates our main point, namely that in the case where gravity remains strong at large $c$, here obtained by assuming the mass to be of order $c^2$, the strong time dilation effects are encoded in the effective non-relativistic theory as (twistless) torsion.

Finally it is interesting to point out that in this second expansion the Newtonian potential is actually zero. In this strong gravitational regime the non-relativistic physics of the Schwarzschild metric can be completely described by non-zero torsion and a curved spatial metric, with a vanishing Newtonian potential!

\section*{Acknowledgements}
It is a pleasure to thank H. Afshar, E. Bergshoeff, J. Hartong, H. Nur, N. Obers, J. Rosseel and \c{C}.Yunus for valuable discussions and correspondence. Part of this work has been previously presented at IPM Tehran, State University of Yerevan and the Simons Center for Geometry and Physics. This work was partially supported by the Bo\u{g}azi\c{c}i University Research Fund under grant number 17B03P1.  

\appendix
\section{Probe Particle Motion: a first analysis}\label{partsec}
The main difference with the previous work of  \cite{Dautcourt:1990, Dautcourt:1996pm, Tichy:2011te} is that we allowed for non-vanishing torsion, which amounts to allowing a divergent term $\os{\Gamma}{-2}_{\m\n}^\l$ in the Levi-Civitta connection. One question this raises is if this divergence is problematic for test particle motion, as it will directly enter the geodesic equations. In this appendix we work out this expansion in some detail and show that a consistent set of equations can be obtained. We are missing however at the moment a good physical intepretation of the different fields and terms appearing in this expansion and equations. This is a very interesting question which we leave for future work. 

\subsection{Setup}
The relativistic equations for the velocity field of a massive particle in a gravitational background can be written as
\begin{equation}
\mathbb{P}^{\m}{}_{\n}V^\r\nabla_{\r}V^\n=0\label{veleq}
\end{equation}
where
\begin{equation}
V^\m=\frac{dX^\m}{d\s}\qquad \mathbb{P}^{\m}{}_{\n}=\delta^{\m}_\n-V^\mu V_\nu/V^2\qquad V^2=g_{\m\n}V^\m V^\n<0
\end{equation}
We choose to work in the manifestly reparametrization invariant formulation, i.e the equations are invariant under $\delta_\l V^\mu=\lambda(x) V^\m$. To connect a solution of the above equations to a particle trajectory $X^\mu(\s)$ one would have to solve the additional equations
\begin{equation}
V^\mu(X(\s))=\frac{d X^\m(\s)}{d\s}\label{trajeq}
\end{equation}
Note that these are invariant under reparametrization symmetry $\s=\sigma'$ via $\lambda(X(\sigma))=\frac{d\sigma}{d\sigma'}$. 

Here we will only attemtp an expansion of the velocity equations, leaving a detailed large $c$ expansion of \eqref{trajeq} for the future.  Our starting point is then an expansion of the form
\begin{equation}
V^\m=\sum_{i=-1}^\infty \os{V}{$i$}^\m c^{-i}\label{vexp}
\end{equation}
Note that the overall power in $c$ of this series can be set to anything by using the rescaling symmetry. We fix this by conventionally choosing the leading power to be -1. This is motivated by the natural choice of parameters where $\s(\tau_\mu X^\mu)$ is a $c$-independent function. The rescaling symmetries preserving this convention and the expansion ansatz \eqref{vexp} are then of the form
\begin{equation}
\lambda=\sum_{i=0}^\infty \os{\l}{$i$} c^{-i}
\end{equation}
Leading to the transformations
\begin{equation}
\delta_\os{\lambda}{$j$}\os{V}{$i$}^\mu=\begin{cases}
\os{\lambda}{$j$}\os{V}{$i$-$j$}\!^\mu&\quad\mbox{if }i\geq j-1\\
0&\quad\mbox{otherwise}
\end{cases}
\end{equation}

\subsection{Leading Order}
We now expand the equations \eqref{veleq} in a background satisfying the TTNC equations of table \ref{eomtable}. It turns out that for this expansion to be consistent, one needs
\begin{equation}
\tau_\r \os{V}{-1}^\r=0
\end{equation}
This can be verified by assuming otherwise and observing that the expansion becomes inconsistent.

First note that for a physical particle (i.e. timelike velocity) we need
\begin{equation}
\os{V}{-2}^2=-(\tau_{\r}\os{V}{0}^\r)^2+\hat{h}_{\r\s}\os{V}{-1}^\r\os{V}{-1}^\s<0\,.
\end{equation}
we then define
\begin{equation}
\os{|V|}{-1}=\sqrt{-\os{V}{-2}^2}\,.
\end{equation}

It is then an interesting observation that when the torsion is twistless one finds that the most singular term of order $c^4$, which is the naive LO, in the velocity equation expansion automatically cancels. Vanishing of the NLO term at order $c^2$ is then equivalent to  
\begin{equation}
\os{V}{-1}^\l \os{\nabla}{nc}_\l \os{V}{-1}^\m=(\os{V}{-1}^\r\partial_\r\log\os{|V|}{-1})\os{V}{-1}^\m+(\tau_\r\os{V}{0}^\r)^2 h^{\m\s}\hat{a}_\s
\end{equation}
Note that before doing the expansion there was the fact that contraction of the relativistic equations \eqref{veleq} with the vector $g_{\m\r}V^\r$ lead to a trivial equality. As the leading term of that vector is $-\tau_\mu \tau_\r \os{V}{0}^\r$ the leading equation in the expansion should trivially vanish when contracted with $\tau_\mu$. Indeed this is the case. Also checking invariance under the different orders of diffeomorphisms is rather straightforward.

It is also easy to verify that the above equation is invariant under both
\begin{equation}
\delta_{\os{\l}{-1}}\os{V}{0}=\os{\l}{0}\os{V}{-1}\,\ \delta_{\os{\l}{0}}\os{V}{0}=\os{\l}{0}\os{V}{0}\qquad\mbox{and}\quad \delta_{\os{\l}{1}}\os{V}{0}=\os{\l}{1}\os{V}{-1}
\end{equation}
Note that via gauge-fixing the $\os{\l}{0}$ invariance we can get rid of $\os{V}{1}^\mu$. One straightforward gauge choice is $\os{|V|}{-1}=1$, which leads to
\begin{equation}
\os{V}{-1}^\l \os{\nabla}{nc}_\l \os{V}{-1}^\m=(1+\hat h_{\r\s}\os{V}{-1}^\r\os{V}{-1}^\s) h^{\m\l}\hat{a}_\l
\end{equation}
This LO equation is important as we see that non-zero torsion forces the velocity coefficient $\os{V}{-1}^\m$ to be non-zero. This is the crucial difference between motion of a particle in a torsional background and one without torsion. If the torsion is zero then one can choose $\os{V}{-1}=0$ and $\os{V}{0}^\mu$, which is the non-relativistic velocity, will satisfy the standard non-relativistic NC geodesic equation. In case the torsion, and hence also $\os{V}{-1}$ is non-zero the equations become much more complicated, as we explore in the next subsection.

\subsection{Higher order}
We can work out the expansion of \eqref{veleq} up to higher orders. A reasonable thing to do to go up to those orders where the equations are fully determined by the background fields appearing in the TTNC equations of table \eqref{eomtable}. It is technically useful to do this expansion when fixing reparametrization invariance. One way to do so is the choice
\begin{eqnarray}
(\tau_\r\os{V}{0}^\r)^2&=&1+\hat{h}_{\r\s}\os{V}{-1}^\r\os{V}{-1}^\s\nonumber\\
\tau_\r\os{V}{0}^\r\tau_\s\os{V}{1}^\s&=&\hat{h}_{\r\s}\os{V}{-1}^\r\os{V}{0}^\s\label{constreqs}\\
2\tau_\r\os{V}{0}^\r\tau_\s\os{V}{2}^\s&=&\hat{h}_{\r\s}(2\os{V}{-1}^\r\os{V}{1}^\s+\os{V}{0}^\r\os{V}{0}^\s)-(\tau_\r \os{V}{1}^\r)^2-2\hat{\Phi}(\tau_\r \os{V}{0}^\r)^2-\hat{\beta}_{\r\s}\os{V}{-1}^\r\os{V}{-1}^\s\nonumber
\end{eqnarray}
which after some calculation leads to the following set of equations for the velocity coefficients
\begin{eqnarray}
\os{V}{-1}^\l \os{\nabla}{nc}_\l \os{V}{-1}^\m&=&(\tau_\r\os{V}{0}^\r)^2 h^{\m\l}\hat{a}_\l\nonumber\\
\os{V}{-1}^\l \os{\nabla}{nc}_\l \os{V}{0}^\m+\os{V}{0}^\l \os{\nabla}{nc}_\l \os{V}{-1}^\m&=&\tau_\r \os{V}{0}^\r\left(\hat a_\sigma \os{V}{-1}^\sigma\hat{\tau}^\mu+2\tau_\s\os{V}{1}^\s h^{\m\l} \hat a_\l\right)\label{expgeo}\\
\os{V}{-1}^\l \os{\nabla}{nc}_\l \os{V}{1}^\m+\os{V}{0}^\l \os{\nabla}{nc}_\l \os{V}{0}^\m+\os{V}{1}^\l \os{\nabla}{nc}_\l \os{V}{-1}^\m&=&\tau_\r\tau_\s\left(\os{V}{1}^\r\os{V}{1}^\s+2\os{V}{0}^\r\os{V}{2}^\s\right)h^{\m\l} \hat a_\l\nonumber\\
&&+\tau_\r\left(\os{V}{0}^\r\os{V}{0}^\s+2\os{V}{1}^{(\r}\os{V}{-1}^{\s)}\right)\left(\tau_\s \hat{\beta}^{\m\l}\hat{a}_\l+\hat a_\s\hat{\tau}^\m\right)\nonumber\\
&&+\os{V}{-1}^\r\os{V}{-1}^\s\left(\frac{1}{2}\hat{h}_{\r\l}\os{\nabla}{nc}_\s\hat{\beta}^{\m\l}-\hat{\tau}^\m\hat{\calk}_{\r\s}\right)\nonumber
\end{eqnarray}
Note that when the torsion is zero, i.e. $\hat{a}_\mu=0$, then one can take $\os{V}{-1}^\m=0$ and the complicated set of equations above reduces to the simple equation $\os{V}{0}^\l \os{\nabla}{nc}_\l \os{V}{0}^\m=0$, which is the geodesic equation in torsionless NC gravity, which is equivalent to Newton's second law in the presence of a gravitational potential. It would be very interesting to understand the physical meaning of all the additional terms that appear when the torsion is turned on.

Note that one can rewrite the equations in manifest reparametrization invariant form by substituting in the above expressions
\begin{equation}
\os{V}{-1}\rightarrow \os{\l}{0}\os{V}{-1}\,,\qquad \os{V}{0}\rightarrow \os{\l}{0}\os{V}{0}+\os{\l}{1}\os{V}{-1}\,,\qquad \os{V}{1}\rightarrow \os{\l}{0}\os{V}{1}+\os{\l}{1}\os{V}{0}+\os{\l}{2}\os{V}{-1}\,,\qquad \os{V}{2}\rightarrow \os{\l}{0}\os{V}{2}+\os{\l}{1}\os{V}{1}+\os{\l}{2}\os{V}{0}+\os{\l}{3}\os{V}{-1}  
\end{equation}
where we should take for the $\lambda$ coefficients the solutions of
\begin{eqnarray*}
\os{\l}{0}&=&\frac{1}{\sqrt{(\tau_\r\os{V}{0}^\r)^2-\hat{h}_{\r\s}\os{V}{-1}^\r\os{V}{-1}^\s}}\\
\frac{\os{\l}{1}}{\os{\l}{0}^3}&=&\tau_\r\os{V}{0}^\r\tau_\s\os{V}{1}^\s-\hat{h}_{\r\s}\os{V}{-1}^\r\os{V}{0}^\s\\
2\frac{\os{\l}{2}}{\os{\l}{0}^3}-3\frac{\os{\l}{1}^2}{\os{\l}{0}^4}&=&2\tau_\r\os{V}{0}^\r\tau_\s\os{V}{2}^\s-\hat{h}_{\r\s}(2\os{V}{-1}^\r\os{V}{1}^\s-\os{V}{0}^\r\os{V}{0}^\s)+(\tau_\r \os{V}{1}^\r)^2+2\hat{\Phi}(\tau_\r \os{V}{0}^\r)^2+\hat{\beta}_{\r\s}\os{V}{-1}^\r\os{V}{-1}^\s
\end{eqnarray*}

\subsection{An example solution}
In section \ref{schsec} we explained how Schwarzschild with a mass of order $c^2$ provides a background that fits into the large $c$ expansion with non-zero torsion. We can now expand the geodesics of this relativistic background along the lines above and this should provide a solution to the equations \eqref{expgeo}. Here we show that this is indeed the case.

The simplest example is that of radially infalling geodesics in Schwarzschild, which are tangent to the vectorfield
\begin{equation}
V^\mu=\lambda(t,r)\left(1,-c\sqrt{\frac{2MG_\mathrm{N}}{r}}\left(1-\frac{2MG_\mathrm{N}}{r}\right),0,0\right)
\end{equation}
From this one can simply read off that
\begin{eqnarray}
\os{V}{-1}^\m&=&\left(0,-\lambda(t,r)\sqrt{\frac{2MG_\mathrm{N}}{r}}\left(1-\frac{2MG_\mathrm{N}}{r}\right),0,0\right)\\
\os{V}{0}^\m&=&\left(\lambda(t,r),0,0,0\right)
\end{eqnarray}
As expected one can check that this provides a solution of \eqref{expgeo}.

\bibliographystyle{utphys}
\bibliography{NC_lit}
\end{document}